\documentclass[aip, jcp, twocolumn, a4paper, floatfix, superscriptaddress, reprint, 10pt]{revtex4-1}  
\usepackage{pdfpages}
\usepackage{pgffor}

\usepackage{graphicx}
\usepackage{xcolor}
\usepackage{amsmath}
\usepackage{amssymb}
\usepackage{xspace}
\usepackage{footnote}
\usepackage{relsize}

\usepackage[paperwidth=210mm,
            paperheight=297mm,
            left=20mm,
            top=10mm,
            textwidth=170mm,
            marginparsep=3mm,
            marginparwidth=30mm,
            textheight=730pt,
            footskip=50pt,portrait]
           {geometry}
           
\makeatletter
\AtBeginDocument{\let\LS@rot\@undefined}
\makeatother

\newcommand{\mE}{\ensuremath{\boldsymbol{\mathcal{E}}}}

\begin{document}

\title{Mean-Field Theory of Water-Water Correlations in Electrolyte Solutions}

\author{David M. Wilkins}

\email{david.wilkins@epfl.ch}

\affiliation{Laboratory of Computational Science and Modeling, IMX, \'Ecole Polytechnique F\'ed\'erale de Lausanne, 1015 Lausanne, Switzerland}
\affiliation{Laboratory for fundamental BioPhotonics, Institutes of Bioengineering and Materials Science and Engineering, School of Engineering, and Lausanne Centre for Ultrafast Science,  \'Ecole Polytechnique F\'ed\'erale de Lausanne, CH-1015 Lausanne, Switzerland}

\author{David E. Manolopoulos}

\affiliation{Physical and Theoretical Chemistry Laboratory, University of Oxford, South Parks Road, Oxford OX1 3QZ, UK}

\author {Sylvie Roke}

\affiliation{Laboratory for fundamental BioPhotonics, Institutes of Bioengineering and Materials Science and Engineering, School of Engineering, and Lausanne Centre for Ultrafast Science,  \'Ecole Polytechnique F\'ed\'erale de Lausanne, CH-1015 Lausanne, Switzerland}

\author{Michele Ceriotti}

\affiliation{Laboratory of Computational Science and Modeling, IMX, \'Ecole Polytechnique F\'ed\'erale de Lausanne, 1015 Lausanne, Switzerland}

\begin{abstract}
Long-range ion induced water-water correlations were recently observed in femtosecond elastic second harmonic scattering experiments of electrolyte solutions. To further the qualitative understanding of these correlations, we derive an analytical expression that quantifies ion induced dipole-dipole correlations in a non-interacting gas of dipoles. This model is a logical extension of Debye-H\"uckel theory that can be used to qualitatively understand how the combined electric field of the ions induces correlations in the orientational distributions of the water molecules in an aqueous solution. The model agrees with results from molecular dynamics simulations and provides an important starting point for further theoretical work.     
\end{abstract}

\maketitle

The electric field of a solvated ion in water induces orientational ordering in the surrounding solvent molecules. 
However, the length scale over which this ordering persists has been a topic of significant debate, at least in part because the range at which correlations can be detected depends on the experimental probe.\cite{Marcus2009} 
The results of neutron diffraction,\cite{Howell1996,Soper2006} X-ray scattering,\cite{Bouazizi2006,Bouazizi2007} 
dielectric relaxation,\cite{Buchner1999} and femtosecond pump-probe experiments,\cite{Omta2003} as well as atomistic simulations of the reorientation timescales of water molecules\cite{Stirnemann2013} and of the vibrational spectrum of
solutions,\cite{Smith2007,Funkner2012} have suggested that the ordering of the surrounding water molecules by ions extends no further than about 3 solvation shells (around 0.8 nm) for sub-molar concentrations. 
On the other hand, infrared photodissociation experiments,\cite{Obrien2010,Obrien2012} and a study combining terahertz and femtosecond infrared spectroscopies,\cite{Tielrooij2010} have found evidence for ordering extended to longer ranges.
Molecular dynamics simulations looking directly at the orientational correlations between water molecules showed that the presence of ionic solutes have an effect on these correlations at distances of more than 1 nm.\cite{Zhang2014}

Femtosecond elastic second harmonic scattering (fs-ESHS)\cite{Shen1989,Roke2012} measurements have recently been used to probe the orientational order of water molecules in H$_{2}$O and D$_{2}$O electrolyte solutions,\cite{Chen2016}
revealing intensity changes that are already detectable at micromolar concentrations, and which are identical for more than 20 different electrolytes. The non-specificity of the fs-ESHS response, its magnitude, and its onset at low concentration point to its long-range origin. The isotope exchange experiment, together with the recorded polarization combinations (in conjunction with the selection rules for nonlinear light scattering experiments \cite{Bersohn1966,Roke2012}) show that the recorded changes in the fs-ESHS response in the concentration range from 1 $\mu$M - 100 mM arise from water-water correlations that are induced by the ions (and not from the ions themselves). This effect shows intriguing correlations with 
changes in the surface tension of dilute electrolyte solutions, suggesting that the same microscopic phenomenon underlying the second-harmonic signal can have an impact on macroscopic observables.

In this Communication we derive an analytical expression for the correlations induced in a non-interacting gas of dipoles by the electric field of ions. This expression is a natural extension to a simple Debye-H\"uckel model,
which has been shown to qualitatively capture the concentration-dependence of the second-harmonic response,\cite{Shelton2009,Chen2016}
and can be used to elucidate the nature, the range and the energetics of the weak ion-induced ordering probed by fs-ESHS.\footnote{The model derived in this paper is an extension of that found in D.M.W.'s D.Phil thesis, University of Oxford, 2016.} The expression provides a benchmark for a fundamental understanding of the interplay of 
ion-dipole and ion-ion interactions.
By comparison with classical molecular dynamics simulations of dilute NaCl solutions, we demonstrate that both of these factors are needed to 
characterize the ion-induced solvent correlations.

We begin by considering the water molecules in an ionic solution to be point dipoles that interact only with the solute, and have no explicit dipole-dipole interactions. Thus,
the orientational ordering of these dipoles is caused only by the electric field due to the ions.
Although this might appear to be a harsh assumption -- and it certainly implies that the model cannot report on \emph{short-ranged} hydrogen-bonding and dipole-dipole interactions -- the
dipolar screening is implicitly included through macroscopic quantities such as the dielectric constant and the local field factor. We will also show later that dipole-dipole interactions can be included in a refined version of the model, and have no impact on the long-range behavior. 

\begin{figure}[!ht]
\centering
\includegraphics[width=8.5cm,keepaspectratio=true]{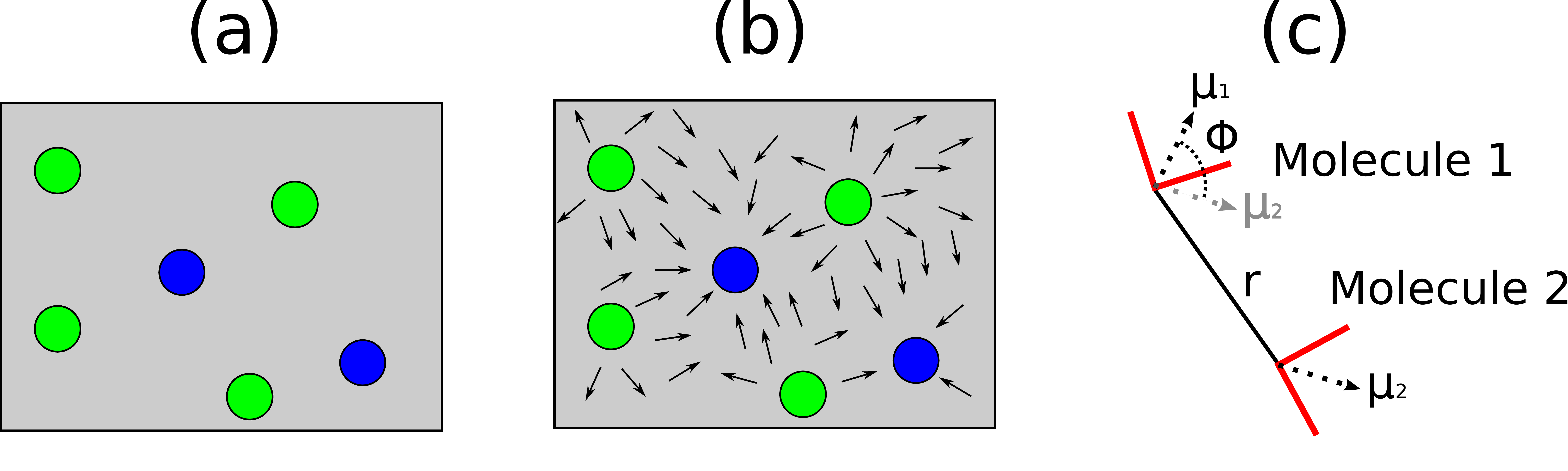}
\caption{\label{fig:mean_field_model} Illustration of the mean-field model considered in this communication: (a) ions are embedded in a dielectric continuum, and then (b) a uniform gas of independent point dipoles is added to the system. (c) Dipole correlation function $\left\langle\cos\phi\right\rangle(r)$: $r$ is the distance between two dipoles, $\boldsymbol{\mu}_{1}$ and $\boldsymbol{\mu}_{2}$ are their dipole moment vectors and $\cos\phi = \hat{\boldsymbol{\mu}}_{1}\cdot\hat{\boldsymbol{\mu}}_{2}$.}
\end{figure}

Figs.~\ref{fig:mean_field_model}(a) and (b) show how this model is built up: firstly, the ions are taken to be point charges in a dielectric continuum, with an appropriate spatial distribution, after which the system is filled with a uniform gas of independent dipoles,\cite{Shelton2009} which will align with the local electric field. 
We then define the dipole correlation function for two solvent molecules separated by a distance $r$ (that is, the average inner product of two dipoles as a function of their separation),
\begin{equation}\label{eq:dipole_correlation_definition}
\left\langle\cos\phi\right\rangle(r) = \frac{1}{V}\int_{V}\left\langle\boldsymbol{\hat{\mu}}(\boldsymbol{R})\cdot\boldsymbol{\hat{\mu}}(\boldsymbol{R} + \boldsymbol{r})\right\rangle_{\text{o+i}}\,{\rm d}^{3} \boldsymbol{R},
\end{equation}

\noindent where $\boldsymbol{\hat{\mu}}(\boldsymbol{R})$ is the unit vector in the direction of the dipole moment of a molecule at $\boldsymbol{R}$, $V$ is the volume of the system and ``o+i'' denotes an average over molecular orientations and ionic positions. Fig.~\ref{fig:mean_field_model}(c) illustrates how the angle $\phi$ is defined for two representative water molecules.

In the Supplementary Information (SI), we show that by taking a Taylor expansion in the reciprocal temperature $\beta = 1/k_{\rm B} T$, we can make the approximation,
\begin{equation} \label{eq:approx_efield}
\left\langle\cos\phi\right\rangle(r) \simeq \frac{1}{V}\left(\frac{\beta\mu}{3}\right)^{2}\int_{V}\left\langle \boldsymbol{E}(\boldsymbol{R})\cdot\boldsymbol{E}(\boldsymbol{R}+\boldsymbol{r})\right\rangle_{\text{i}}\,{\rm d}^{3} \boldsymbol{R},
\end{equation}
where $\boldsymbol{E}(\boldsymbol{R})$ is the total electric field at position $\boldsymbol{R}$ due to all of the ions in the solution, and $\mu$ is the permanent dipole moment of a water molecule.
The subscript ``i'' indicates that the average is taken over the positions of ions. For simplicity of notation, any angular brackets in the following work without a subscript are taken over ion positions only. Eqn.~\eqref{eq:approx_efield} shows that in our model the correlation between dipoles is proportional to the correlation between electric fields, which are taken to be the only source of ordering for the molecules.

The electric field $\boldsymbol{E}(\boldsymbol{R})$ at a given position is the sum of electric fields due to all of the ions. This allows us to write,
\begin{equation}
\boldsymbol{E}(\boldsymbol{R}) = \sum_{m\in\text{ions}} \frac{e Z_{m} f_{0}}{4\pi\epsilon_{0}\epsilon_{\rm r}} \boldsymbol{\mathcal{E}}(\boldsymbol{R}-\boldsymbol{r}_{m}),
\end{equation}
with $Z_{m}$ the charge of the $m^{\rm th}$ ion in units of the electron charge $e$ and $\boldsymbol{r}_{m}$ the position of this ion, $f_{0}$ the Onsager local field factor,\cite{Onsager1936} $\epsilon_{0}$ the vacuum permittivity, $\epsilon_{\rm r}$ the solvent dielectric constant, and $\boldsymbol{\mathcal{E}}(\boldsymbol{r})$ the electric field associated with individual ions (most commonly the Coulomb field, $\boldsymbol{r}/{r^{3}}$). This gives
\begin{multline}\label{eq:multiple_fields}
\left\langle\cos\phi\right\rangle(r) \simeq \\ \frac{A}{V}  \sum_{m,n} Z_{m}Z_{n} \int_{V} \left\langle \boldsymbol{\mathcal{E}}(\boldsymbol{R} - \boldsymbol{r}_{m})\cdot\boldsymbol{\mathcal{E}}(\boldsymbol{R} + \boldsymbol{r} - \boldsymbol{r}_{n})\right\rangle\,{\rm d}^{3}\boldsymbol{R},
\end{multline}

\noindent in which we have defined $A = \left(\frac{\beta\mu f_{0} e}{12 \pi \epsilon_{0}\epsilon_{\rm r}}\right)^{2}$.

In the thermodynamic ($V\rightarrow\infty$) limit the integral in Eqn.~\eqref{eq:multiple_fields} is taken over all space and can be most conveniently expressed in reciprocal space,
\begin{multline}
\left\langle\cos\phi\right\rangle(r) \simeq \\ \frac{A}{V} \int \boldsymbol{\mathcal{E}}(\boldsymbol{K})\cdot\boldsymbol{\mathcal{E}}(-\boldsymbol{K}) \left\langle \sum_{m,n} Z_{m} Z_{n} e^{i\boldsymbol{K}\cdot\left(\boldsymbol{r}_{m} - \boldsymbol{r}_{n}\right)}\right\rangle e^{i\boldsymbol{K}\cdot{\boldsymbol{r}}}\,\frac{{\rm d}^{3}\boldsymbol{K}} {\left(2\pi\right)^{3}},
\end{multline}

\noindent where $\boldsymbol{\mathcal{E}}(\boldsymbol{K})$ is the Fourier transform of the field function $\boldsymbol{\mathcal{E}}(\boldsymbol{r})$. The term in angular brackets is proportional to the charge-charge structure factor $S(\boldsymbol{K})$ of the ions.\cite{Barrat2003} This gives the dipole correlation function in terms of the ion number density $\rho$ as,
\begin{multline}\label{eq:fields_structure_factor}
\left\langle\cos\phi\right\rangle(r) \simeq \\
\frac{\rho}{\left(2\pi\right)^{3}} \left(\frac{\beta \mu e f_{0}}{12 \pi \epsilon_{0} \epsilon_{\rm r}}\right)^{2} \int \left| \boldsymbol{\mathcal{E}}(\boldsymbol{K})\right|^{2} S(\boldsymbol{K}) e^{i \boldsymbol{K}\cdot\boldsymbol{r}}\,{\rm d}^{3}\boldsymbol{K}.
\end{multline}

\begin{figure}[!ht]
\centering
\includegraphics[width=8.5cm,keepaspectratio=true]{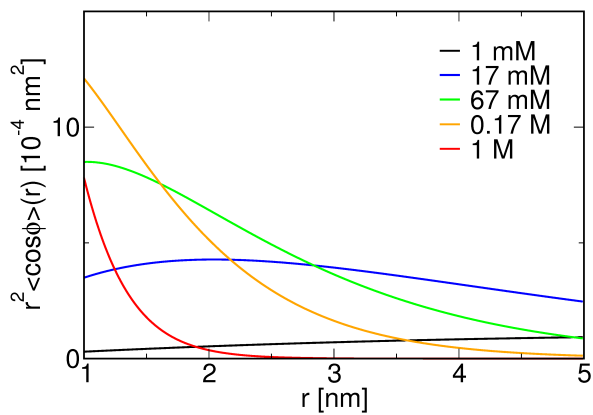}
\caption{\label{fig:compare_sim_theory} Solvent dipole-dipole correlation functions predicted by Eqn.~\eqref{eq:DH_corr_function} at different concentrations. 
}
\end{figure}

The most appropriate mean-field model can be obtained by taking the field function $\mE(\boldsymbol{r})$ to be the Coulomb field $\boldsymbol{r} / r^{3}$ (corresponding to $\mE(\boldsymbol{K}) = -4 \pi i \boldsymbol{K} / K^{2}$), and 
using the Debye-H\"uckel (DH) structure factor \cite{Barrat2003} $S(K) = \frac{2 K^{2}}{K^{2} + \kappa^{2}}$, where $\kappa = \left(\frac{2 \rho \beta Z^{2} e^{2}}{\epsilon_{0}\epsilon_{\rm r}}\right)^{1/2}$ is the inverse Debye length. This gives
\begin{equation}\label{eq:DH_corr_function}
\left\langle\cos\phi\right\rangle_{\text{\tiny{DH}}}(r) = \frac{\rho}{2\pi}\left(\frac{\beta\mu e f_{0}}{3\epsilon_{0}\epsilon_{\rm r}}\right)^{2} \frac{e^{-\kappa r}}{r}.
\end{equation}
The variation of $\left\langle\cos\phi\right\rangle_{\text{\tiny{DH}}}(r)$ with ion concentration is instructive. As seen in Figure~\ref{fig:compare_sim_theory}, 
for small $\rho$, an increase in concentration leads to an increase in correlation between solvent dipoles, while for large $\rho$ the $e^{-\kappa r}$ factor dominates. Increasing the concentration results in ions being more screened and with a lesser propensity to orient solvent dipoles.
It should be also noted that, at all of the concentrations shown in Fig.~\ref{fig:compare_sim_theory}, the dipolar correlations at distances above 5 nm are very small. However, because the number of water molecules further than 5 nm away is very large,  these correlations can be measured by fs-ESHS experiments, a testimony to the exquisite sensitivity of the probe. 

Eqn.~\eqref{eq:fields_structure_factor} allows us to investigate the interplay between 
the ion-ion spatial correlations (encoded in $S(\boldsymbol{K})$) 
and the ion-dipole orientational correlations (due to the electric field $\mE(\boldsymbol{r})$).
By changing the form of  $S(\boldsymbol{K})$, one can estimate the response to an arbitrary distribution of ions:
for instance, one could extend this model to investigate the correlations induced by charges on an interface.
A particularly instructive example involves a completely uncorrelated arrangement of ions. 
This random-ion (RI) model is equivalent to setting $S(\boldsymbol{K}) = 2$, 
which leads to dipole-dipole correlations corresponding to Eqn.~\eqref{eq:DH_corr_function} with $\kappa = 0$, while the concentration $\rho$ is kept constant.
At all concentrations, this RI model leads to
increased dipole-dipole correlations, because of the 
lack of screening of the Coulombic ion-dipole 
interaction by the correlated cloud of counterions.
It is worth stressing that, although it might be 
appealing to qualitatively discuss the dampening of
correlations in terms of the exponentially-screened 
DH field of an ion, this is not an appropriate model.
Such a screened-field/random ions (SF-RI) model 
amounts to setting $\boldsymbol{\mathcal{E}}(\boldsymbol{r}) =-\nabla\,\frac{e^{-\kappa r}}{r} = \boldsymbol{r}\left(\frac{e^{-\kappa r}}{r^{3}} + \frac{\kappa e^{-\kappa r}}{r^{2}}\right)$ and
$S(\boldsymbol{K}) = 2$. The resulting functional form
of the induced dipole-dipole correlations resembles
that of the full DH model at short distances, but then
leads to unphysical anticorrelations at large distance
(see the SI).

\begin{figure}[t]
\centering
\includegraphics[width=8.5cm,keepaspectratio=true]{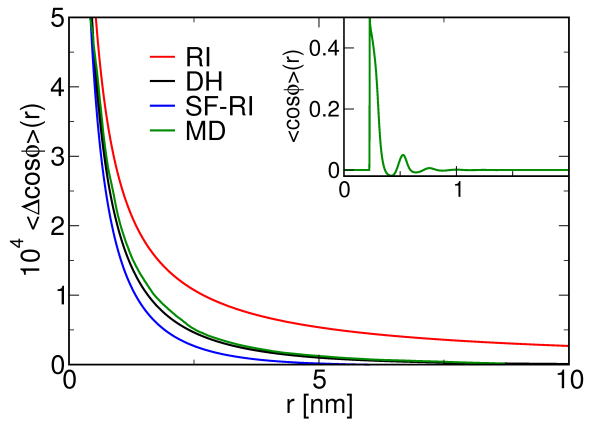}
\caption{\label{fig:three_cases} Comparison of the solvent dipole correlation function for the full Debye-H\"uckel theory (black line), the random-ion approximation (red line) 
and the screened-field plus random-ion approximation (blue line), with a salt concentration of 8 mM at $T = 300$ K.
We also show the correlation function calculated from MD at this concentration (green line).
In all cases we have
subtracted the correlation function for pure water
at the same temperature. Note that this correlation
is zero for the mean-field model, and so 
$\left\langle\Delta\cos\phi\right\rangle(r)=\left\langle\cos\phi\right\rangle(r)$ for all curves but MD.
Inset: the absolute correlation function $\left\langle\cos\phi\right\rangle(r)$ calculated from MD, showing considerable structure at short range.}
\end{figure}

Fig.~\ref{fig:three_cases} compares the predicted $\left\langle\cos\phi\right\rangle(r)$ using the full DH theory, the RI and the SF-RI models, and the correlations computed from a MD simulation using a $\sim$ 20 nm cubic box with about 264,000 TIP4P/2005 water molecules.\cite{Chen2016}
All curves correspond to a salt concentration of 8 mM and a temperature of 300 K. The other physical constants used are described in the SI. 
Comparison with MD results in Fig.~\ref{fig:three_cases} shows that
only the full DH model captures the correct long-range behavior of the dipole-dipole correlations -- although the short range structure is clearly absent.
Neglecting ion-ion spatial correlations artificially increases the orientational correlations, since randomly distributed ions cannot efficiently screen the fields of other ions.
A picture in which one interprets dipole-dipole correlations in terms of the screened electrostatic field of the ions, while providing a qualitative 
picture of the physics, is inconsistent with the 
linearized-Boltzmann structure of the mean-field model, and fails to quantitatively reproduce the MD results. %
This comparison demonstrates that the long-ranged dipole-dipole correlations are most naturally interpreted as being due to the bare electric field of the ions. The correlations are modulated by short range interactions (which are not included in this model), and by the presence of ion-ion spatial correlations, which result in the screening of the Coulomb field. This latter effect leads to decreased dipole-dipole correlations and provides an explanation for the saturation of the fs-ESHS signal at high electrolyte concentrations.

We note that the mean-field model 
can be further improved to include more physical effects.
$\left\langle\cos\phi\right\rangle_{\text{\tiny{DH}}}(r)$ diverges in the $r\rightarrow 0$ limit because of 
the singularity in the electric field at the ion positions.
It is possible to remove this short-distance divergence by restricting the volume of space in which water molecules can be found; however, the fact that two water molecules have a distance of minimum approach, below which $\left\langle\cos\phi\right\rangle(r)$ is not meaningful, makes the divergence irrelevant.
We can also estimate the impact of neglecting dipole-dipole interactions, by re-introducing them in a perturbative fashion.
This can be done by following the procedure used to derive the approximation in Eqn.~\eqref{eq:approx_efield}, including also the dipole-dipole interaction. In doing so, we find (as described in the SI) that the lowest-order term in $\left\langle\cos\phi\right\rangle(r)$ that includes the dipole-dipole forces is proportional to $\beta^{4} e^{-\kappa r}/r^{7}$. This term decays much more rapidly than does the model of Eqn.~\eqref{eq:DH_corr_function}, and makes essentially no contribution at long enough distances: above 0.33 nm, the magnitude of this correction is less than 1 \% of the magnitude of $\left\langle\cos\phi\right\rangle_{\text{\tiny{DH}}}(r)$, and less than $10^{-3}$ \% above 1 nm.

The computed residual orientational correlation of dipoles at a distance of several nm is extremely small, but since it involves many dipoles the total change in free energy may be non-negligible.  
In order to elucidate the free energy scale associated with ion-induced long-range dipole-dipole correlations, 
we evaluate the total energetic contribution associated with the oriented dipoles at distances larger than a chosen cut-off length $r_\text{c}$, which reads (see the SI),~\cite{Hill1986}%
\begin{equation}
U = 4\pi\rho_{\rm S}\mu\int_{r_{\rm c}}^{\infty} r^{2} E(r) \mathcal{L}\left(\beta\mu E(r)\right) \,{\rm d} r,
\end{equation}
where $\mathcal{L}(x) = \coth(x) - 1/x$ is the Langevin function and $\rho_{\rm S}$ is the solvent density. The mean electric field $E(r)$ around an ion is given by Debye-H\"uckel theory. The integral can be computed by expanding the integrand as a Taylor series in $\beta$.

\begin{figure}[t]
\centering
\includegraphics[width=8.5cm,keepaspectratio=true]{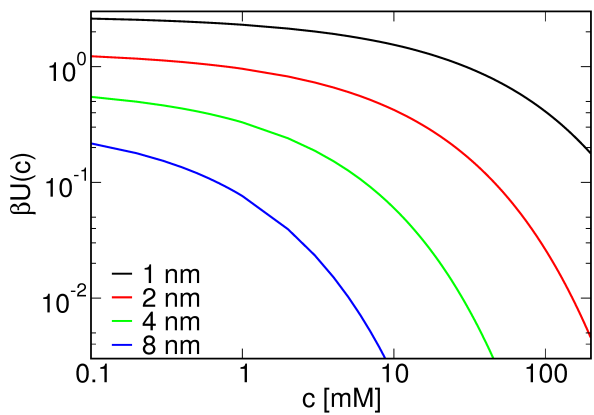}
\caption{\label{fig:dipole_field_energy} Energy of the dipoles oriented by a single ion as a function of ion concentration, for varying short-range cutoff distances $r_{\rm c}$.
}
\end{figure}

Fig.~\ref{fig:dipole_field_energy} shows the total energetic contribution of the dipoles oriented by an ion as a function of the electrolyte concentration and for different cut-off distances. 
At mM concentrations, dipolar order beyond the Bjerrum length ($\sim 0.7$ nm in water at 300 K) is associated with an energy scale of about 3 $k_B T$, and even the tails beyond 4 nm correspond to a significant fraction of $k_B T$.
Due to the large number of dipoles in the far region, the \emph{collective} effect is significant even though each ion-dipole interaction is very small.
Thus, it is plausible that ion-induced dipole-dipole correlations extending well beyond the Bjerrum length could lead to measurable changes in the macroscopic energy (as observed in the surface tension measurements of Ref.~\onlinecite{Chen2016}). As this analysis is performed with a very simplified model, this conclusion is not definitive, and a more quantitative analysis should include changes in the long-range dipole-dipole order in the bulk and in the surface region. These changes could then be connected to changes in the free energy of the surface and the bulk region. 

In conclusion, we have shown that long-range, non-specific electrolyte-induced correlations in water as recently observed in fs-ESHS experiments can be captured by a simple mean-field model that treats water molecules as non-interacting dipoles oriented by the  electrostatic field of ions, which are themselves correlated following Debye-H\"uckel theory. 
Although one can intuitively understand the 
orientational correlations as arising from the exponentially-screened field of correlated ions,
a more accurate picture, leading to quantitative
predictions of MD simulations, regards 
them as arising from unscreened ion-dipole 
correlations that combine destructively 
when the physically relevant ion-ion correlations are included.
This model is very useful to pinpoint what we think is the main physical origin of the electrolyte-induced change in the fs-ESHS intensity and to estimate the length and energy scale of the effect. 
It does not, however, explain the dramatic isotope effects that are seen in experiments,\cite{Chen2016} or the temperature dependence of the fs-ESHS signal. 
As such it is clearly only a first step in a complete description of the experimental data, which should also include a re-evaluation of the molecular hyperpolarizability tensor,\cite{Tocci2016} particularly when probed by femtosecond laser pulses.\cite{Liang2017}

\section*{Supplementary Information}

See supplementary information for more detailed derivations of the formulas used in the main text, as well as a list of the numerical values of physical constants used.

\begin{acknowledgments}

The authors thank Damien Laage for helpful discussions, and Halil Okur and Yixing Chen for critical reading of the manuscript. D.M.W. and M.C. acknowledge funding from the Swiss National Science Foundation (Project ID 200021\_163210). S. R. acknowledges funding from the Julia Jacobi Foundation and the European Research Council (grant number 616305).

\end{acknowledgments}

\foreach \x in {1,2,3,4,5,6,7,8,9}
{%
\clearpage


\section*{Supplementary material}

## I. HIGH-TEMPERATURE APPROXIMATION TO DIPOLE-DIPOLE CORRELATION FUNCTION

Eqn. (1) of the main text contains the expression,

$$\langle \hat{\boldsymbol{\mu}}(\boldsymbol{R}) \cdot \hat{\boldsymbol{\mu}}(\boldsymbol{R} + \boldsymbol{r}) \rangle_{\mathrm{o+i}}, \tag{S1}$$

where "o+i" represents an average over the orientations of the dipole at position $\boldsymbol{R}_1 = \boldsymbol{R}$ and the dipole at position $\boldsymbol{R}_2 = \boldsymbol{R} + \boldsymbol{r}$, and over the positions of ions. We wish to convert this into an average only over the ion positions. To do so we note that,

$$\langle \hat{\boldsymbol{\mu}}(\boldsymbol{R}_1) \cdot \hat{\boldsymbol{\mu}}(\boldsymbol{R}_2) \rangle_{\mathrm{o+i}} = \left\langle \frac{\int \mathrm{d}\Omega_{\boldsymbol{R}_1} \int \mathrm{d}\Omega_{\boldsymbol{R}_2} \, \hat{\boldsymbol{\mu}}(\boldsymbol{R}_1) \cdot \hat{\boldsymbol{\mu}}(\boldsymbol{R}_2) e^{-\beta H(\Omega_{\boldsymbol{R}_1}, \Omega_{\boldsymbol{R}_2})}}{\int \mathrm{d}\Omega_{\boldsymbol{R}_1} \int \mathrm{d}\Omega_{\boldsymbol{R}_2} \, e^{-\beta H(\Omega_{\boldsymbol{R}_1}, \Omega_{\boldsymbol{R}_2})}} \right\rangle_{\mathrm{i}}, \tag{S2}$$

with the subscript "i" indicating an average only over ion positions, $\Omega_{\boldsymbol{R}_1} = (\cos\theta_1, \phi_1)$ describing the orientation of the dipole at $\boldsymbol{R}_1$ and $\Omega_{\boldsymbol{R}_2} = (\cos\theta_2, \phi_2)$ the orientation of the dipole at $\boldsymbol{R}_2$. $H(\Omega_{\boldsymbol{R}_1}, \Omega_{\boldsymbol{R}_2})$ is the energy of the pair of dipoles as a function of the orientations.

If the dipoles interact only with their local electric field then the energy of the pair of molecules is,

$$H(\Omega_{\boldsymbol{R}_1}, \Omega_{\boldsymbol{R}_2}) = -\mu \left( \hat{\boldsymbol{\mu}}(\boldsymbol{R}_1) \cdot \boldsymbol{E}(\boldsymbol{R}_1) + \hat{\boldsymbol{\mu}}(\boldsymbol{R}_2) \cdot \boldsymbol{E}(\boldsymbol{R}_2) \right), \tag{S3}$$

where $\mu$ is the permanent dipole moment of a molecule and $\boldsymbol{E}(\boldsymbol{R}_i)$ the electric field due to all ions at $\boldsymbol{R}_i$.

Writing $\hat{\boldsymbol{\mu}}(\boldsymbol{R}_1)$ in terms of an orthonormal basis,

$$\hat{\boldsymbol{\mu}}(\boldsymbol{R}_1) = \boldsymbol{e}_{\boldsymbol{R}_1}^1 \cos\theta_1 + \boldsymbol{e}_{\boldsymbol{R}_1}^2 \sin\theta_1 \cos\phi_1 + \boldsymbol{e}_{\boldsymbol{R}_1}^3 \sin\theta_1 \sin\phi_1, \tag{S4}$$

with $\boldsymbol{e}_{\boldsymbol{R}_1}^1$ the unit vector in the direction of the electric field at $\boldsymbol{R}_1$ and $\boldsymbol{e}_{\boldsymbol{R}_1}^2$ and $\boldsymbol{e}_{\boldsymbol{R}_1}^3$ two vectors orthogonal to the field. A similar expression can be given for $\hat{\boldsymbol{\mu}}(\boldsymbol{R}_2)$ and we find that,

$$H(\Omega_{\boldsymbol{R}_1}, \Omega_{\boldsymbol{R}_2}) = -\mu \left( E(\boldsymbol{R}_1) \cos\theta_1 + E(\boldsymbol{R}_2) \cos\theta_2 \right). \tag{S5}$$

Having written the unit vectors in terms of the angles that specify the orientations of the molecules, we can carry out the integrations in Eqn. (S2) to give,

$$\langle \hat{\boldsymbol{\mu}}(\boldsymbol{R}_1) \cdot \hat{\boldsymbol{\mu}}(\boldsymbol{R}_2) \rangle_{\mathrm{o+i}} = \left\langle \mathcal{L}\left[ \beta\mu E(\boldsymbol{R}_1) \right] \boldsymbol{e}_{\boldsymbol{R}_1}^1 \cdot \boldsymbol{e}_{\boldsymbol{R}_2}^1 \mathcal{L}\left[ \beta\mu E(\boldsymbol{R}_2) \right] \right\rangle_{\mathrm{i}}. \tag{S6}$$

Here, $\mathcal{L}[x] = \coth(x) - 1/x$ is the Langevin function. Taking a Taylor series of this expression we find that the lowest-order approximation is,

$$\langle \hat{\boldsymbol{\mu}}(\boldsymbol{R}_1) \cdot \hat{\boldsymbol{\mu}}(\boldsymbol{R}_2) \rangle_{\mathrm{o+i}} \simeq \left( \frac{\beta\mu}{3} \right)^2 \langle \boldsymbol{E}(\boldsymbol{R}_1) \cdot \boldsymbol{E}(\boldsymbol{R}_2) \rangle_{\mathrm{i}}. \tag{S7}$$

This allows Eqn. (1) of the main text to be rewritten as,

$$\langle \cos\phi \rangle (r) \simeq \frac{1}{V} \int_V \left( \frac{\beta\mu}{3} \right)^2 \langle \boldsymbol{E}(\boldsymbol{R}) \cdot \boldsymbol{E}(\boldsymbol{R} + \boldsymbol{r}) \rangle \, \mathrm{d}^3\boldsymbol{r}, \tag{S8}$$

with the subscript "i" dropped for simplicity of notation.

## II. PHYSICAL CONSTANTS USED

Table I describes the physical constants used in this Communication.

TABLE I. The physical properties used in this letter.

| Electronic properties of water | | | |
|---|---|---|---|
| Property | Value | Ref. | Notes |
| Solvent dipole moment ($\mu$) | 2.305 D | S1 | – |
| Solvent local field factor ($f_0$) | 3.15 | S2 | A local field factor has not yet been calculated for the TIP4P/2005 water molecule. |
| Solvent dielectric constant ($\varepsilon_r$) | 60 | S1 | – |
| Thermodynamic properties of simulations | | | |
| Property | Value | Ref. | Notes |
| Temperature (T) | 300 K | – | – |
| Solvent number density ($\rho_S$) | $3.337 \times 10^{-2}$ Å$^{-3}$ | – | – |

## III. SCREENED-FIELD APPROXIMATION

In the screened-field, random-ion (SF/RI) model of the main text, the electric field around each ion is given by the screened Debye-Hückel expression,

$$\mathcal{E}(\boldsymbol{r}) = -\nabla \frac{e^{-\kappa r}}{r} \tag{S9}$$

$$= \left( \frac{e^{-\kappa r}}{r^2} + \frac{\kappa e^{-\kappa r}}{r} \right) \frac{\boldsymbol{r}}{r}, \tag{S10}$$

and the structure factor is given by $S(\boldsymbol{K}) = 2$. The Fourier transform of the field function is,

$$\mathcal{E}(\boldsymbol{K}) = -4\pi i \frac{\boldsymbol{K}}{K^2 + \kappa^2}. \tag{S11}$$

The correlation function of Eqn. (6) in the main text is given by,

$$
\begin{aligned}
\langle \cos\phi \rangle \, (r) &\simeq \frac{\rho}{(2\pi)^3} \left( \frac{\beta \mu e f_0}{12\pi \epsilon_0 \epsilon_r} \right)^2 \int |\mathcal{E}(\boldsymbol{K})|^2 S(\boldsymbol{K}) e^{i\boldsymbol{K}\cdot\boldsymbol{r}} \, \mathrm{d}^3\boldsymbol{K}, \\
&= \frac{4\rho}{\pi} \left( \frac{\beta \mu e f_0}{12\pi \epsilon_0 \epsilon_r} \right)^2 \int \frac{K^2 e^{i\boldsymbol{K}\cdot\boldsymbol{r}}}{(K^2 + \kappa^2)^2} \, \mathrm{d}^3\boldsymbol{K}, \\
&= \frac{\rho}{\pi} \left( \frac{\beta \mu e f_0}{6 \epsilon_0 \epsilon_r} \right)^2 \frac{e^{-\kappa r} \, (\kappa r - 2)}{r}.
\end{aligned}
\tag{S12}
$$

Fig. S1 shows the predictions of Eqn. (S12) for the same concentrations as considered in Fig. 2 of the main text. The SF-RI model predicts correlations that are qualitatively different to those predicted by the DH model (and thus to those observed in simulations), with unphysical anticorrelations appearing at long distances.

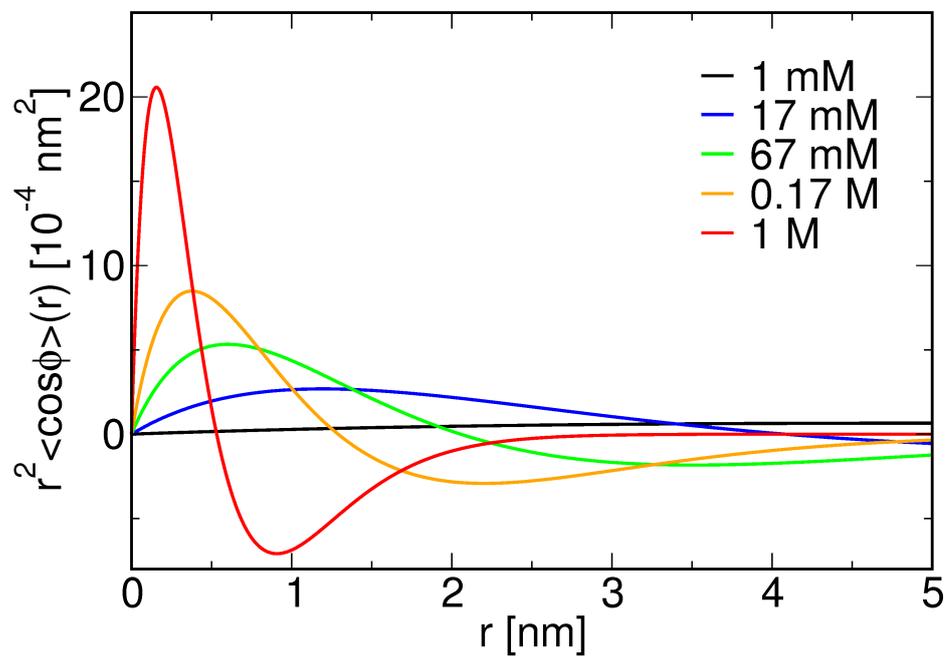

FIG. S1. Solvent dipole-dipole correlation functions calculated with the results of Eqn. (S12) at different concentrations.

## IV. SHORT-RANGE DIVERGENCE OF THE CORRELATION FUNCTION

Eqn. (7) in the main text gives an expression for the correlation function $\langle \cos\phi \rangle (r)$ within Debye-Hückel (DH) theory. However, this equation has the apparent drawback that at $r \to 0$, $\langle \cos\phi \rangle (r)$ diverges as $1/r$. The reason for this can be seen quite straightforwardly: according to Eqn. (2), $\langle \cos\phi \rangle (r \to 0) \sim \langle \boldsymbol{E}(0) \cdot \boldsymbol{E}(0) \rangle = \langle E^2 \rangle$, the mean-squared electric field experienced by a water molecule.

In Eqn. (1) the correlation function is defined as an integral over all space,

$$\langle \cos\phi \rangle (r) = \frac{1}{V} \int_V \langle \hat{\boldsymbol{\mu}}(\boldsymbol{R}) \cdot \hat{\boldsymbol{\mu}}(\boldsymbol{R} + \boldsymbol{r}) \rangle_{\text{o+i}} \, \mathrm{d}^3 \boldsymbol{R}, \tag{1}$$

which includes volume elements in which a molecule is at the position of an ion and $E^2$ is infinite. This means that $\langle E^2 \rangle$ will also diverge.

We note that it is possible to eliminate this short-distance divergence of the correlation function by restricting the volume of space over which the integral is taken, in such a way that only the short-distance behaviour is significantly affected.[S3] However, even without using this excluded-volume correction, this divergence does not represent a disadvantage of the model: two water molecules cannot approach each other closer than a hard-sphere radius of around 0.25 nm, below which $\langle \Delta \cos\phi \rangle (r) = 0$. We do not implement this hard-sphere radius correction in this communication for the sake of simplicity, as we are interested only in the long-distance behaviour of the water-water orientational correlations.

## V. DIPOLE-DIPOLE INTERACTIONS

In a true aqueous system, the water molecules are mutually oriented due not only to the electric fields at their positions, but also to dipole-dipole interactions between them. To account for this interaction, the energy of Eqn. (S3) must be replaced by,

$$H(\Omega_{\boldsymbol{R}_1}, \Omega_{\boldsymbol{R}_2}, \boldsymbol{r}) = -\mu \left( \hat{\boldsymbol{\mu}}(\boldsymbol{R}_1) \cdot \boldsymbol{E}(\boldsymbol{R}_1) + \hat{\boldsymbol{\mu}}(\boldsymbol{R}_2) \cdot \boldsymbol{E}(\boldsymbol{R}_2) \right) + \frac{\alpha}{r^3} \left( \hat{\boldsymbol{\mu}}(\boldsymbol{R}_1) \cdot \hat{\boldsymbol{\mu}}(\boldsymbol{R}_2) - 3 \left[ \hat{\boldsymbol{\mu}}(\boldsymbol{R}_1) \cdot \hat{\boldsymbol{r}} \right] \left[ \hat{\boldsymbol{r}} \cdot \hat{\boldsymbol{\mu}}(\boldsymbol{R}_2) \right] \right), \tag{S13}$$

with $\alpha = \mu^2 f_0^2 / 4\pi\varepsilon_0\varepsilon_r$. Eqn. (S2) becomes,

$$\langle \hat{\boldsymbol{\mu}}(\boldsymbol{R}_1) \cdot \hat{\boldsymbol{\mu}}(\boldsymbol{R}_2) \rangle_{\text{o+i}} = \left\langle \frac{\int \mathrm{d}\Omega_{\boldsymbol{R}_1} \int \mathrm{d}\Omega_{\boldsymbol{R}_2} \, \hat{\boldsymbol{\mu}}(\boldsymbol{R}_1) \cdot \hat{\boldsymbol{\mu}}(\boldsymbol{R}_2) e^{-\beta H(\Omega_{\boldsymbol{R}_1}, \Omega_{\boldsymbol{R}_2}, \boldsymbol{r})}}{\int \mathrm{d}\Omega_{\boldsymbol{R}_1} \int \mathrm{d}\Omega_{\boldsymbol{R}_2} \, e^{-\beta H(\Omega_{\boldsymbol{R}_1}, \Omega_{\boldsymbol{R}_2}, \boldsymbol{r})}} \right\rangle_{\text{i}}. \tag{S14}$$

Unlike Eqn. (S2), Eqn. (S14) cannot be evaluated analytically before making an approximation, so we instead use a Taylor expansion in $\beta$ before integrating over orientations. We write,

$$\hat{\boldsymbol{\mu}}(\boldsymbol{R}_1) = \hat{\boldsymbol{r}} \cos\theta_1 + \hat{\boldsymbol{x}} \sin\theta_1 \cos\phi_1 + \hat{\boldsymbol{y}} \sin\theta_1 \sin\phi_1, \tag{S15a}$$

$$\hat{\boldsymbol{\mu}}(\boldsymbol{R}_2) = \hat{\boldsymbol{r}} \cos\theta_2 + \hat{\boldsymbol{x}} \sin\theta_2 \cos\phi_2 + \hat{\boldsymbol{y}} \sin\theta_2 \sin\phi_2, \tag{S15b}$$

where $\hat{\boldsymbol{x}}$, $\hat{\boldsymbol{y}}$ and $\hat{\boldsymbol{r}}$ form an orthonormal set of Cartesian axes. With $\boldsymbol{\Omega_{R_1}} = (\cos\theta_1, \phi_1)$ and $\boldsymbol{\Omega_{R_2}} = (\cos\theta_2, \phi_2)$ we write the total energy as,

$$H(\Omega_{\boldsymbol{R}_1}, \Omega_{\boldsymbol{R}_2}, \boldsymbol{r}) = -\mu \left[ \cos\theta_1 e_r(\boldsymbol{R}_1) + \sin\theta_1 \cos\phi_1 e_x(\boldsymbol{R}_1) + \sin\theta_1 \sin\phi_1 e_y(\boldsymbol{R}_1) + \right.$$
$$\cos\theta_2 e_r(\boldsymbol{R}_2) + \sin\theta_2 \cos\phi_2 e_x(\boldsymbol{R}_2) + \sin\theta_2 \sin\phi_2 e_y(\boldsymbol{R}_2) \left] + \right.$$
$$\frac{\alpha}{r^3} \left[ \sin\theta_1 \cos\phi_1 \sin\theta_2 \cos\phi_2 + \sin\theta_1 \sin\phi_1 \sin\theta_2 \sin\phi_2 - 2\cos\theta_1 \cos\theta_2 \right], \tag{S16}$$

with the definitions,

$$e_x(\boldsymbol{R}_i) = \boldsymbol{E}(\boldsymbol{R}_i) \cdot \boldsymbol{x}, \tag{S17a}$$

$$e_y(\boldsymbol{R}_i) = \boldsymbol{E}(\boldsymbol{R}_i) \cdot \boldsymbol{y}, \tag{S17b}$$

$$e_r(\boldsymbol{R}_i) = \boldsymbol{E}(\boldsymbol{R}_i) \cdot \boldsymbol{r}. \tag{S17c}$$

Using the fact that $\hat{\boldsymbol{\mu}}(\boldsymbol{R}_1) \cdot \hat{\boldsymbol{\mu}}(\boldsymbol{R}_2) = \cos\theta_1 \cos\theta_2 + \sin\theta_1 \cos\phi_1 \sin\theta_2 \cos\phi_2 + \sin\theta_1 \sin\phi_1 \sin\theta_2 \sin\phi_2$, we expand the Boltzmann factors in both the numerator and the denominator of Eqn. (S14) up to order $\beta^4$, before carrying out the integrals over $\boldsymbol{\Omega}_{\boldsymbol{R}_1}$ and $\boldsymbol{\Omega}_{\boldsymbol{R}_2}$. The quantity in angular brackets is then expanded up to order $\beta^4$. Using the following symmetry relations,

$$\langle [\boldsymbol{E}(\boldsymbol{R}_i) \cdot \hat{\boldsymbol{x}}] \, [\hat{\boldsymbol{x}} \cdot \boldsymbol{E}(\boldsymbol{R}_i)] \rangle = \langle [\boldsymbol{E}(\boldsymbol{R}_i) \cdot \hat{\boldsymbol{y}}] \, [\hat{\boldsymbol{y}} \cdot \boldsymbol{E}(\boldsymbol{R}_i)] \rangle = \langle [\boldsymbol{E}(\boldsymbol{R}_i) \cdot \hat{\boldsymbol{r}}] \, [\hat{\boldsymbol{r}} \cdot \boldsymbol{E}(\boldsymbol{R}_i)] \rangle = \frac{1}{3} \, \langle \boldsymbol{E}(\boldsymbol{R}_i) \cdot \boldsymbol{E}(\boldsymbol{R}_i) \rangle, \tag{S18a}$$

$$\langle \boldsymbol{E}(\boldsymbol{R}_i) \cdot \boldsymbol{E}(\boldsymbol{R}_i) \rangle = \langle E^2 \rangle, \tag{S18b}$$

$$\langle \, E(\boldsymbol{R}_1)^2 \boldsymbol{E}(\boldsymbol{R}_1) \cdot \boldsymbol{E}(\boldsymbol{R}_2) \rangle = \langle \, E(\boldsymbol{R}_2)^2 \boldsymbol{E}(\boldsymbol{R}_1) \cdot \boldsymbol{E}(\boldsymbol{R}_2) \rangle, \tag{S18c}$$

we obtain,

$$\langle \hat{\boldsymbol{\mu}}(\boldsymbol{R}_1) \cdot \hat{\boldsymbol{\mu}}(\boldsymbol{R}_2) \rangle = \left( \frac{\beta\mu}{3} \right)^2 \langle \boldsymbol{E}(\boldsymbol{R}_1) \cdot \boldsymbol{E}(\boldsymbol{R}_2) \rangle + \frac{2\alpha^3\beta^3}{75 r^9} - \frac{2}{135} \, (\beta\mu)^4 \, \langle E(\boldsymbol{R}_1)^2 \boldsymbol{E}(\boldsymbol{R}_1) \cdot \boldsymbol{E}(\boldsymbol{R}_2) \rangle + \frac{2\alpha^2\mu^2}{675} \frac{\beta^4}{r^6} \, \langle \boldsymbol{E}(\boldsymbol{R}_1) \cdot \boldsymbol{E}(\boldsymbol{R}_2) \rangle + \mathcal{O} \left( \beta^5 \right). \tag{S19}$$

The first term in Eqn. (S19) is the same as that derived in Section I, while the second term is due only to the dipole-dipole interaction. This term is also present in neat water, and so does not describe a correlation induced by the presence of ions. After taking a Taylor series of Eqn. (S14) in $\beta$, there are also terms in $1/r^3$ and $1/r^6$, but by symmetry the integral of these terms over all orientations gives zero (so that the $1/r^9$ term is the lowest-order one due to dipole-dipole interactions). The third term is independent of the dipole-dipole interactions (and will be present even when these interactions are not), and the fourth term represents the lowest-order correction to the correlation function due to the presence of dipole-dipole interactions. The correction to $\langle \cos\phi \rangle_{\text{DH}}(r)$ due to this term is,

$$\langle \cos\phi \rangle_{\text{DH}}(r) = \frac{\rho}{75\pi} \left( \frac{\beta^2 \mu Z e f_0 \alpha}{3\varepsilon_0 \varepsilon_{\text{r}}} \right)^2 \frac{e^{-\kappa r}}{r^7}. \tag{S20}$$

Fig. S2 shows the ratio $\langle \Delta \cos\phi \rangle_{\text{DH}}(r) / \langle \cos\phi \rangle_{\text{DH}}(r)$ as a function of $r$, where we see that for $r \geq 0.33$ nm, the correction term has a magnitude less than 1 % of the true correlation function. This distance corresponds approximately to the boundary of the first solvation shell. This result suggests that the neglect of the dipole-dipole interactions is justified in the derivation of this model.

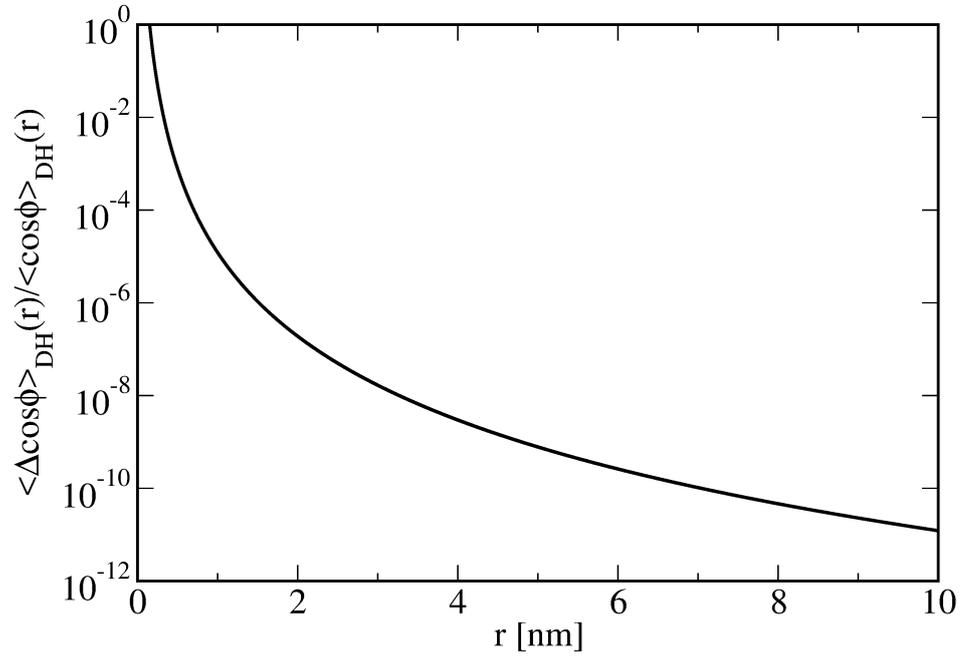

FIG. S2. Magnitude of the correction term of Eqn. (S20) relative to the Debye-Hückel correlation function $\langle \cos \phi \rangle_{\mathrm{DH}}(r)$.

## VI. ENERGY STORED IN THE DIPOLE FIELD

The energy of a water molecule with dipole moment $\boldsymbol{\mu}$, a distance $r$ from an ion is $U = -\mu E(r) \cos\theta$, where $E(r)$ is the electric field due to the ion at the position of the water molecule and $\cos\theta$ is the projection of the dipole moment on the vector from the ion to the molecule. The mean energy of such a water molecule (averaged over all possible orientations) is then,[S4]

$$
\begin{aligned}
\langle U \rangle (r) &= \frac{\int_0^\pi e^{-\beta U(r)} U(r) \sin\theta \, \mathrm{d}\theta}{\int_0^\pi e^{-\beta U(r)} \sin\theta \, \mathrm{d}\theta}, \\
&= \frac{\int_{-1}^1 e^{-\beta \mu E(r) \cos\theta} \cos\theta \, \mathrm{d}(\cos\theta)}{\int_{-1}^1 e^{-\beta \mu E(r) \cos\theta} \, \mathrm{d}(\cos\theta)} \left( -E(r)\mu \right), \\
&= \mu E(r) \mathcal{L}\left( \beta \mu E(r) \right),
\end{aligned}
\tag{S21}
$$

with $\mathcal{L}(x) = \coth(x) - 1/x$ the Langevin function. Since we are considering a single ion screening by all other ions, the appropriate electric field to use is the Debye-Hückel field,

$$
E(r) = \frac{Zef_0}{4\pi\epsilon_0 \epsilon_{\mathrm{r}}} \left( \frac{e^{-\kappa r}}{r^2} + \frac{\kappa e^{-\kappa r}}{r} \right).
\tag{S22}
$$

The total energy of water molecules at least a distance $r_{\mathrm{c}}$ from an ion is given by integrating Eqn. (S21) over the corresponding region of space,

$$
U = \rho_{\mathrm{S}} \int_0^{2\pi} \mathrm{d}\phi \int_{-1}^1 \mathrm{d}(\cos\theta) \int_{r_{\mathrm{c}}}^\infty \mathrm{d}r \, r^2 \langle U \rangle (r),
\tag{S23}
$$

which when integrated over polar angles gives Eqn. (8) in the main text. In order to compute this integral, we took a Taylor series of the integrand (with converged results given by truncating the series at $\mathcal{O}\left( \beta^8 \right)$).


}


\begin{thebibliography}{25}%
\makeatletter
\providecommand \@ifxundefined [1]{%
 \@ifx{#1\undefined}
}%
\providecommand \@ifnum [1]{%
 \ifnum #1\expandafter \@firstoftwo
 \else \expandafter \@secondoftwo
 \fi
}%
\providecommand \@ifx [1]{%
 \ifx #1\expandafter \@firstoftwo
 \else \expandafter \@secondoftwo
 \fi
}%
\providecommand \natexlab [1]{#1}%
\providecommand \enquote  [1]{``#1''}%
\providecommand \bibnamefont  [1]{#1}%
\providecommand \bibfnamefont [1]{#1}%
\providecommand \citenamefont [1]{#1}%
\providecommand \href@noop [0]{\@secondoftwo}%
\providecommand \href [0]{\begingroup \@sanitize@url \@href}%
\providecommand \@href[1]{\@@startlink{#1}\@@href}%
\providecommand \@@href[1]{\endgroup#1\@@endlink}%
\providecommand \@sanitize@url [0]{\catcode `\\12\catcode `\$12\catcode
  `\&12\catcode `\#12\catcode `\^12\catcode `\_12\catcode `\%12\relax}%
\providecommand \@@startlink[1]{}%
\providecommand \@@endlink[0]{}%
\providecommand \url  [0]{\begingroup\@sanitize@url \@url }%
\providecommand \@url [1]{\endgroup\@href {#1}{\urlprefix }}%
\providecommand \urlprefix  [0]{URL }%
\providecommand \Eprint [0]{\href }%
\providecommand \doibase [0]{http://dx.doi.org/}%
\providecommand \selectlanguage [0]{\@gobble}%
\providecommand \bibinfo  [0]{\@secondoftwo}%
\providecommand \bibfield  [0]{\@secondoftwo}%
\providecommand \translation [1]{[#1]}%
\providecommand \BibitemOpen [0]{}%
\providecommand \bibitemStop [0]{}%
\providecommand \bibitemNoStop [0]{.\EOS\space}%
\providecommand \EOS [0]{\spacefactor3000\relax}%
\providecommand \BibitemShut  [1]{\csname bibitem#1\endcsname}%
\let\auto@bib@innerbib\@empty
%</preamble>
\bibitem [{\citenamefont {Marcus}(2009)}]{Marcus2009}%
  \BibitemOpen
  \bibfield  {author} {\bibinfo {author} {\bibfnamefont {Y.}~\bibnamefont
  {Marcus}},\ }\href@noop {} {\bibfield  {journal} {\bibinfo  {journal} {Chem.
  Rev.}\ }\textbf {\bibinfo {volume} {109}},\ \bibinfo {pages} {1346} (\bibinfo
  {year} {2009})}\BibitemShut {NoStop}%
\bibitem [{\citenamefont {Howell}\ and\ \citenamefont
  {Neilson}(1996)}]{Howell1996}%
  \BibitemOpen
  \bibfield  {author} {\bibinfo {author} {\bibfnamefont {I.}~\bibnamefont
  {Howell}}\ and\ \bibinfo {author} {\bibfnamefont {G.~W.}\ \bibnamefont
  {Neilson}},\ }\href@noop {} {\bibfield  {journal} {\bibinfo  {journal} {J.
  Phys. Condens. Matt.}\ }\textbf {\bibinfo {volume} {8}},\ \bibinfo {pages}
  {4455} (\bibinfo {year} {1996})}\BibitemShut {NoStop}%
\bibitem [{\citenamefont {Soper}\ and\ \citenamefont
  {Weckstr{\"o}m}(2006)}]{Soper2006}%
  \BibitemOpen
  \bibfield  {author} {\bibinfo {author} {\bibfnamefont {A.~K.}\ \bibnamefont
  {Soper}}\ and\ \bibinfo {author} {\bibfnamefont {K.}~\bibnamefont
  {Weckstr{\"o}m}},\ }\href@noop {} {\bibfield  {journal} {\bibinfo  {journal}
  {Biophys. Chem.}\ }\textbf {\bibinfo {volume} {124}},\ \bibinfo {pages} {180}
  (\bibinfo {year} {2006})}\BibitemShut {NoStop}%
\bibitem [{\citenamefont {Bouazizi}\ \emph {et~al.}(2006)\citenamefont
  {Bouazizi}, \citenamefont {Nasr}, \citenamefont {Ja{\r{i}}dane},\ and\
  \citenamefont {Bellissent-Funel}}]{Bouazizi2006}%
  \BibitemOpen
  \bibfield  {author} {\bibinfo {author} {\bibfnamefont {S.}~\bibnamefont
  {Bouazizi}}, \bibinfo {author} {\bibfnamefont {S.}~\bibnamefont {Nasr}},
  \bibinfo {author} {\bibfnamefont {N.}~\bibnamefont {Ja{\r{i}}dane}}, \ and\
  \bibinfo {author} {\bibfnamefont {M.-C.}\ \bibnamefont {Bellissent-Funel}},\
  }\href@noop {} {\bibfield  {journal} {\bibinfo  {journal} {J. Phys. Chem. B}\
  }\textbf {\bibinfo {volume} {110}},\ \bibinfo {pages} {23515} (\bibinfo
  {year} {2006})}\BibitemShut {NoStop}%
\bibitem [{\citenamefont {Bouazizi}\ and\ \citenamefont
  {Nasr}(2007)}]{Bouazizi2007}%
  \BibitemOpen
  \bibfield  {author} {\bibinfo {author} {\bibfnamefont {S.}~\bibnamefont
  {Bouazizi}}\ and\ \bibinfo {author} {\bibfnamefont {S.}~\bibnamefont
  {Nasr}},\ }\href@noop {} {\bibfield  {journal} {\bibinfo  {journal} {J. Mol.
  Struct.}\ }\textbf {\bibinfo {volume} {837}},\ \bibinfo {pages} {206}
  (\bibinfo {year} {2007})}\BibitemShut {NoStop}%
\bibitem [{\citenamefont {Buchner}, \citenamefont {Hefter},\ and\ \citenamefont
  {May}(1999)}]{Buchner1999}%
  \BibitemOpen
  \bibfield  {author} {\bibinfo {author} {\bibfnamefont {R.}~\bibnamefont
  {Buchner}}, \bibinfo {author} {\bibfnamefont {G.~T.}\ \bibnamefont {Hefter}},
  \ and\ \bibinfo {author} {\bibfnamefont {P.~M.}\ \bibnamefont {May}},\
  }\href@noop {} {\bibfield  {journal} {\bibinfo  {journal} {J. Phys. Chem. A}\
  }\textbf {\bibinfo {volume} {103}},\ \bibinfo {pages} {1} (\bibinfo {year}
  {1999})}\BibitemShut {NoStop}%
\bibitem [{\citenamefont {Omta}\ \emph {et~al.}(2003)\citenamefont {Omta},
  \citenamefont {Kropman}, \citenamefont {Woutersen},\ and\ \citenamefont
  {Bakker}}]{Omta2003}%
  \BibitemOpen
  \bibfield  {author} {\bibinfo {author} {\bibfnamefont {A.~W.}\ \bibnamefont
  {Omta}}, \bibinfo {author} {\bibfnamefont {M.~F.}\ \bibnamefont {Kropman}},
  \bibinfo {author} {\bibfnamefont {S.}~\bibnamefont {Woutersen}}, \ and\
  \bibinfo {author} {\bibfnamefont {H.~J.}\ \bibnamefont {Bakker}},\
  }\href@noop {} {\bibfield  {journal} {\bibinfo  {journal} {Science}\ }\textbf
  {\bibinfo {volume} {301}},\ \bibinfo {pages} {347} (\bibinfo {year}
  {2003})}\BibitemShut {NoStop}%
\bibitem [{\citenamefont {Stirnemann}\ \emph {et~al.}(2013)\citenamefont
  {Stirnemann}, \citenamefont {Wernersson}, \citenamefont {Jungwirth},\ and\
  \citenamefont {Laage}}]{Stirnemann2013}%
  \BibitemOpen
  \bibfield  {author} {\bibinfo {author} {\bibfnamefont {G.}~\bibnamefont
  {Stirnemann}}, \bibinfo {author} {\bibfnamefont {E.}~\bibnamefont
  {Wernersson}}, \bibinfo {author} {\bibfnamefont {P.}~\bibnamefont
  {Jungwirth}}, \ and\ \bibinfo {author} {\bibfnamefont {D.}~\bibnamefont
  {Laage}},\ }\href@noop {} {\bibfield  {journal} {\bibinfo  {journal} {J. Am.
  Chem. Soc.}\ }\textbf {\bibinfo {volume} {135}},\ \bibinfo {pages} {11824}
  (\bibinfo {year} {2013})}\BibitemShut {NoStop}%
\bibitem [{\citenamefont {Smith}, \citenamefont {Saykally},\ and\ \citenamefont
  {Geissler}(2007)}]{Smith2007}%
  \BibitemOpen
  \bibfield  {author} {\bibinfo {author} {\bibfnamefont {J.~D.}\ \bibnamefont
  {Smith}}, \bibinfo {author} {\bibfnamefont {R.~J.}\ \bibnamefont {Saykally}},
  \ and\ \bibinfo {author} {\bibfnamefont {P.~L.}\ \bibnamefont {Geissler}},\
  }\href@noop {} {\bibfield  {journal} {\bibinfo  {journal} {J. Am. Chem.
  Soc.}\ }\textbf {\bibinfo {volume} {129}},\ \bibinfo {pages} {13847}
  (\bibinfo {year} {2007})}\BibitemShut {NoStop}%
\bibitem [{\citenamefont {Funkner}\ \emph {et~al.}(2012)\citenamefont
  {Funkner}, \citenamefont {Niehues}, \citenamefont {Schmidt}, \citenamefont
  {Heyden}, \citenamefont {Schwaab}, \citenamefont {Callahan}, \citenamefont
  {Tobias},\ and\ \citenamefont {Havenith}}]{Funkner2012}%
  \BibitemOpen
  \bibfield  {author} {\bibinfo {author} {\bibfnamefont {S.}~\bibnamefont
  {Funkner}}, \bibinfo {author} {\bibfnamefont {G.}~\bibnamefont {Niehues}},
  \bibinfo {author} {\bibfnamefont {D.~A.}\ \bibnamefont {Schmidt}}, \bibinfo
  {author} {\bibfnamefont {M.}~\bibnamefont {Heyden}}, \bibinfo {author}
  {\bibfnamefont {G.}~\bibnamefont {Schwaab}}, \bibinfo {author} {\bibfnamefont
  {K.~M.}\ \bibnamefont {Callahan}}, \bibinfo {author} {\bibfnamefont {D.~J.}\
  \bibnamefont {Tobias}}, \ and\ \bibinfo {author} {\bibfnamefont
  {M.}~\bibnamefont {Havenith}},\ }\href@noop {} {\bibfield  {journal}
  {\bibinfo  {journal} {J. Am. Chem. Soc.}\ }\textbf {\bibinfo {volume}
  {134}},\ \bibinfo {pages} {1030} (\bibinfo {year} {2012})}\BibitemShut
  {NoStop}%
\bibitem [{\citenamefont {O'Brien}\ \emph {et~al.}(2010)\citenamefont
  {O'Brien}, \citenamefont {Prell}, \citenamefont {Bush},\ and\ \citenamefont
  {Williams}}]{Obrien2010}%
  \BibitemOpen
  \bibfield  {author} {\bibinfo {author} {\bibfnamefont {J.~T.}\ \bibnamefont
  {O'Brien}}, \bibinfo {author} {\bibfnamefont {J.~S.}\ \bibnamefont {Prell}},
  \bibinfo {author} {\bibfnamefont {M.~F.}\ \bibnamefont {Bush}}, \ and\
  \bibinfo {author} {\bibfnamefont {E.~R.}\ \bibnamefont {Williams}},\
  }\href@noop {} {\bibfield  {journal} {\bibinfo  {journal} {J. Am. Chem.
  Soc.}\ }\textbf {\bibinfo {volume} {132}},\ \bibinfo {pages} {8248} (\bibinfo
  {year} {2010})}\BibitemShut {NoStop}%
\bibitem [{\citenamefont {O'Brien}\ and\ \citenamefont
  {Williams}(2012)}]{Obrien2012}%
  \BibitemOpen
  \bibfield  {author} {\bibinfo {author} {\bibfnamefont {J.~T.}\ \bibnamefont
  {O'Brien}}\ and\ \bibinfo {author} {\bibfnamefont {E.~R.}\ \bibnamefont
  {Williams}},\ }\href@noop {} {\bibfield  {journal} {\bibinfo  {journal} {J.
  Am. Chem. Soc.}\ }\textbf {\bibinfo {volume} {134}},\ \bibinfo {pages}
  {10228} (\bibinfo {year} {2012})}\BibitemShut {NoStop}%
\bibitem [{\citenamefont {Tielrooij}\ \emph {et~al.}(2010)\citenamefont
  {Tielrooij}, \citenamefont {Garcia-Araez}, \citenamefont {Bonn},\ and\
  \citenamefont {Bakker}}]{Tielrooij2010}%
  \BibitemOpen
  \bibfield  {author} {\bibinfo {author} {\bibfnamefont {K.~J.}\ \bibnamefont
  {Tielrooij}}, \bibinfo {author} {\bibfnamefont {N.}~\bibnamefont
  {Garcia-Araez}}, \bibinfo {author} {\bibfnamefont {M.}~\bibnamefont {Bonn}},
  \ and\ \bibinfo {author} {\bibfnamefont {H.~J.}\ \bibnamefont {Bakker}},\
  }\href@noop {} {\bibfield  {journal} {\bibinfo  {journal} {Science}\ }\textbf
  {\bibinfo {volume} {328}},\ \bibinfo {pages} {1006} (\bibinfo {year}
  {2010})}\BibitemShut {NoStop}%
\bibitem [{\citenamefont {Zhang}\ and\ \citenamefont
  {Galli}(2014)}]{Zhang2014}%
  \BibitemOpen
  \bibfield  {author} {\bibinfo {author} {\bibfnamefont {C.}~\bibnamefont
  {Zhang}}\ and\ \bibinfo {author} {\bibfnamefont {G.}~\bibnamefont {Galli}},\
  }\href@noop {} {\bibfield  {journal} {\bibinfo  {journal} {J. Chem. Phys.}\
  }\textbf {\bibinfo {volume} {141}},\ \bibinfo {pages} {084504} (\bibinfo
  {year} {2014})}\BibitemShut {NoStop}%
\bibitem [{\citenamefont {Shen}(1989)}]{Shen1989}%
  \BibitemOpen
  \bibfield  {author} {\bibinfo {author} {\bibfnamefont {Y.~R.}\ \bibnamefont
  {Shen}},\ }\href@noop {} {\bibfield  {journal} {\bibinfo  {journal} {Annu.
  Rev. Phys. Chem.}\ }\textbf {\bibinfo {volume} {40}},\ \bibinfo {pages} {327}
  (\bibinfo {year} {1989})}\BibitemShut {NoStop}%
\bibitem [{\citenamefont {Roke}\ and\ \citenamefont
  {Gonella}(2012)}]{Roke2012}%
  \BibitemOpen
  \bibfield  {author} {\bibinfo {author} {\bibfnamefont {S.}~\bibnamefont
  {Roke}}\ and\ \bibinfo {author} {\bibfnamefont {G.}~\bibnamefont {Gonella}},\
  }\href@noop {} {\bibfield  {journal} {\bibinfo  {journal} {Annu. Rev. Phys.
  Chem.}\ }\textbf {\bibinfo {volume} {63}},\ \bibinfo {pages} {353} (\bibinfo
  {year} {2012})}\BibitemShut {NoStop}%
\bibitem [{\citenamefont {Chen}\ \emph {et~al.}(2016)\citenamefont {Chen},
  \citenamefont {Okur}, \citenamefont {Gomopoulos}, \citenamefont
  {Macias-Romero}, \citenamefont {Cremer}, \citenamefont {Petersen},
  \citenamefont {Tocci}, \citenamefont {Wilkins}, \citenamefont {Liang},
  \citenamefont {Ceriotti},\ and\ \citenamefont {Roke}}]{Chen2016}%
  \BibitemOpen
  \bibfield  {author} {\bibinfo {author} {\bibfnamefont {Y.}~\bibnamefont
  {Chen}}, \bibinfo {author} {\bibfnamefont {H.~I.}\ \bibnamefont {Okur}},
  \bibinfo {author} {\bibfnamefont {N.}~\bibnamefont {Gomopoulos}}, \bibinfo
  {author} {\bibfnamefont {C.}~\bibnamefont {Macias-Romero}}, \bibinfo {author}
  {\bibfnamefont {P.~S.}\ \bibnamefont {Cremer}}, \bibinfo {author}
  {\bibfnamefont {P.~B.}\ \bibnamefont {Petersen}}, \bibinfo {author}
  {\bibfnamefont {G.}~\bibnamefont {Tocci}}, \bibinfo {author} {\bibfnamefont
  {D.~M.}\ \bibnamefont {Wilkins}}, \bibinfo {author} {\bibfnamefont
  {C.}~\bibnamefont {Liang}}, \bibinfo {author} {\bibfnamefont
  {M.}~\bibnamefont {Ceriotti}}, \ and\ \bibinfo {author} {\bibfnamefont
  {S.}~\bibnamefont {Roke}},\ }\href@noop {} {\bibfield  {journal} {\bibinfo
  {journal} {Sci. Adv.}\ }\textbf {\bibinfo {volume} {2}},\ \bibinfo {pages}
  {e1501891} (\bibinfo {year} {2016})}\BibitemShut {NoStop}%
\bibitem [{\citenamefont {Bersohn}, \citenamefont {Pao},\ and\ \citenamefont
  {Frisch}(1966)}]{Bersohn1966}%
  \BibitemOpen
  \bibfield  {author} {\bibinfo {author} {\bibfnamefont {R.}~\bibnamefont
  {Bersohn}}, \bibinfo {author} {\bibfnamefont {Y.}~\bibnamefont {Pao}}, \ and\
  \bibinfo {author} {\bibfnamefont {H.~L.}\ \bibnamefont {Frisch}},\
  }\href@noop {} {\bibfield  {journal} {\bibinfo  {journal} {J. Chem. Phys.}\
  }\textbf {\bibinfo {volume} {45}},\ \bibinfo {pages} {3184} (\bibinfo {year}
  {1966})}\BibitemShut {NoStop}%
\bibitem [{\citenamefont {Shelton}(2009)}]{Shelton2009}%
  \BibitemOpen
  \bibfield  {author} {\bibinfo {author} {\bibfnamefont {D.~P.}\ \bibnamefont
  {Shelton}},\ }\href@noop {} {\bibfield  {journal} {\bibinfo  {journal} {J.
  Chem. Phys.}\ }\textbf {\bibinfo {volume} {130}},\ \bibinfo {pages} {114501}
  (\bibinfo {year} {2009})}\BibitemShut {NoStop}%
\bibitem [{Note1()}]{Note1}%
  \BibitemOpen
  \bibinfo {note} {The model derived in this paper is an extension of that
  found in D.M.W.'s D.Phil thesis, University of Oxford, 2016.}\BibitemShut
  {Stop}%
\bibitem [{\citenamefont {Onsager}(1936)}]{Onsager1936}%
  \BibitemOpen
  \bibfield  {author} {\bibinfo {author} {\bibfnamefont {L.}~\bibnamefont
  {Onsager}},\ }\href@noop {} {\bibfield  {journal} {\bibinfo  {journal} {J.
  Am. Chem. Soc.}\ }\textbf {\bibinfo {volume} {58}},\ \bibinfo {pages} {1486}
  (\bibinfo {year} {1936})}\BibitemShut {NoStop}%
\bibitem [{\citenamefont {Barrat}\ and\ \citenamefont
  {Hansen}(2003)}]{Barrat2003}%
  \BibitemOpen
  \bibfield  {author} {\bibinfo {author} {\bibfnamefont {J.}~\bibnamefont
  {Barrat}}\ and\ \bibinfo {author} {\bibfnamefont {J.}~\bibnamefont
  {Hansen}},\ }\href@noop {} {\emph {\bibinfo {title} {{Basic Concepts for
  Simple and Complex Liquids}}}}\ (\bibinfo  {publisher} {Cambridge University
  Press},\ \bibinfo {address} {Cambridge},\ \bibinfo {year} {2003})\BibitemShut
  {NoStop}%
\bibitem [{\citenamefont {Hill}(1986)}]{Hill1986}%
  \BibitemOpen
  \bibfield  {author} {\bibinfo {author} {\bibfnamefont {T.~L.}\ \bibnamefont
  {Hill}},\ }\href@noop {} {\emph {\bibinfo {title} {An Introduction to
  Statistical Thermodynamics}}},\ \bibinfo {edition} {2nd}\ ed.\ (\bibinfo
  {address} {New York},\ \bibinfo {year} {1986})\BibitemShut {NoStop}%
\bibitem [{\citenamefont {Tocci}\ \emph {et~al.}(2016)\citenamefont {Tocci},
  \citenamefont {Liang}, \citenamefont {Wilkins}, \citenamefont {Roke},\ and\
  \citenamefont {Ceriotti}}]{Tocci2016}%
  \BibitemOpen
  \bibfield  {author} {\bibinfo {author} {\bibfnamefont {G.}~\bibnamefont
  {Tocci}}, \bibinfo {author} {\bibfnamefont {C.}~\bibnamefont {Liang}},
  \bibinfo {author} {\bibfnamefont {D.~M.}\ \bibnamefont {Wilkins}}, \bibinfo
  {author} {\bibfnamefont {S.}~\bibnamefont {Roke}}, \ and\ \bibinfo {author}
  {\bibfnamefont {M.}~\bibnamefont {Ceriotti}},\ }\href@noop {} {\bibfield
  {journal} {\bibinfo  {journal} {J. Phys. Chem. Lett.}\ }\textbf {\bibinfo
  {volume} {7}},\ \bibinfo {pages} {4311} (\bibinfo {year} {2016})}\BibitemShut
  {NoStop}%
\bibitem [{\citenamefont {Liang}\ \emph {et~al.}(2017)\citenamefont {Liang},
  \citenamefont {Tocci}, \citenamefont {Wilkins}, \citenamefont {Grisafi},
  \citenamefont {Roke},\ and\ \citenamefont {Ceriotti}}]{Liang2017}%
  \BibitemOpen
  \bibfield  {author} {\bibinfo {author} {\bibfnamefont {C.}~\bibnamefont
  {Liang}}, \bibinfo {author} {\bibfnamefont {G.}~\bibnamefont {Tocci}},
  \bibinfo {author} {\bibfnamefont {D.~M.}\ \bibnamefont {Wilkins}}, \bibinfo
  {author} {\bibfnamefont {A.}~\bibnamefont {Grisafi}}, \bibinfo {author}
  {\bibfnamefont {S.}~\bibnamefont {Roke}}, \ and\ \bibinfo {author}
  {\bibfnamefont {M.}~\bibnamefont {Ceriotti}},\ }\href@noop {} {\bibfield
  {journal} {\bibinfo  {journal} {Submitted}\ } (\bibinfo {year}
  {2017})}\BibitemShut {NoStop}%
\end{thebibliography}
\end{document}